\documentclass[%
 reprint,
 amsmath,amssymb,
 aps,
]{revtex4-2}
\usepackage{amsfonts}

\usepackage{graphicx}% Include figure files
\usepackage{subcaption}

\usepackage{dcolumn}% Align table columns on decimal point
\usepackage{braket}
\usepackage{bm}% bold math
\newcommand{\tr}{\text{Tr}}

\begin{document}

\preprint{APS/123-QED}

\title{Entanglement Harvesting of Inertially Moving Unruh-DeWitt Detectors in Minkowski Spacetime}

\author{C. Suryaatmadja}
\email{csuryaat@uwaterloo.ca}
\affiliation{Dept. of Physics \& Astronomy, University of Waterloo, 200 University Avenue, Canada}
\author{W. Cong}%
\affiliation{Faculty of Physics, University of Vienna, Vienna, Austria}%
\author{R.B. Mann}
\affiliation{Dept. of Physics \& Astronomy, University of Waterloo, 200 University Avenue, Canada}%

\date{\today}% It is always \today, today,

\begin{abstract}
We investigate the effects
of relative motion on entanglement harvesting by considering a pair of Unruh-Dewitt detectors moving at arbitrary, but independent and constant velocities, both linearly interacting with the vacuum scalar field. 
Working within the weak coupling approximation, we find that the  Negativity is a (non-elementary) function of the relative velocity of the detectors, as well as their energy gaps and minimal separation.
We find parameter regions where Negativity increases with velocity 
up to a maximum and then decreases, reaching zero at some sublight velocity. At any given relative velocity, the harvested entanglement is inversely correlated with the detector energy gap (at sufficiently high values) and the  distance of closest approach of two detectors.
\end{abstract}

\maketitle

%\tableofcontents

\section{Introduction}

Entanglement harvesting \cite{Valentini1991nonlocalcorr,reznik2003entanglement, Steeg2009,Salton:2014jaa} is the process of entangling detectors through their independent coupling with a field by extracting its vacuum entanglement.
The phenomenon can be described using a quantum multi-partite system in the combined Hilbert space of the detectors and the   field. The field, typically taken to be a scalar field for demonstrative purposes, allows for the transfer of virtual particles between detectors, allowing the otherwise independent detectors to become entangled with each other. This phenomenon, originally studied in flat spacetime \cite{Valentini1991nonlocalcorr,reznik2003entanglement}, has now been investigated in detail in  various contexts, such as
cosmological settings \cite{Steeg2009,Martin-Martinez:2012chf}, non-inertial motion \cite{Salton:2014jaa,Liu:2021dnl}, black hole spacetimes \cite{Gallock-Yoshimura:2021yok,Tjoa:2020eqh}, in the presence of gravitational waves \cite{Gray:2021dfk,Xu:2020pbj} and for detectors in superpositions 
of temporal order \cite{Henderson:2020zax}
and of 
trajectories \cite{Foo:2021gkl}.

A popular detector model used in these studies is the Unruh-DeWitt (UDW) detector \cite{PhysRevD.14.870,DeWitt:1980hx}, a qubit whose states are separated by an energy gap. Historically the UDW detector was used to operationally demonstrate the Unruh and Hawking effects \cite{PhysRevD.14.870,Unruh:1983ms,PhysRevD.13.2188,Israel:1976ur} — the detector response thermalizes when accelerated or placed in a black hole spacetime, with the temperature being proportional to the acceleration and black hole temperature respectively. In recent years, this detector model has gained popularity in the field of relativistic quantum information due to its ability to provide operational information about quantum fields in curved spacetimes, bypassing the usual difficulties associated with field measurements \cite{Sorkin:1993gg,Benincasa:2012rb,Bostelmann:2020unl,Dragan:2012hy}.

The ability to induce or manipulate entanglement between detectors has many promising applications including, but not limited to, curvature measurement \cite{Mart_n_Mart_nez_2012,Ahmadzadegan:2013iua}, quantum energy teleportation \cite{Hotta_2014} and data compression \cite{https://doi.org/10.48550/arxiv.quant-ph/0510031}. The implementation of entanglement harvesting is a valuable addition to the methods available to perform these tasks and possesses numerous advantages over other techniques. These include the ability to entangle detectors without direct interaction between them, generation of entanglement over large distances (including space-like separated regions) and the possibility of exploiting the vacuum as a resource of correlations.

The property of UDWs in the static limit have been well-explored in the past: it is known how entanglement harvesting depends on detector parameters such as their energy gaps and  separation \cite{Pozas2015}, the state of the quantum field \cite{Simidzija:2018ddw} and the presence of boundaries \cite{Liu:2020jaj,Cong:2018vqx,Cong:2020nec}.  For simplicity the detectors are generally taken to be identical.  However it was recently shown \cite{Hu:2022nxc} that
a differential energy gap between the detectors can, under the right circumstances, both extend the harvesting-achievable separation between the detectors and extract more entanglement from the vacuum state than is possible if the detector gaps are identical.

Likewise, little attention has been paid to the effects on entanglement harvesting of the relative inertial motion between detectors. While it is always possible to work in the rest frame of any given detector via a Lorentz transformation, it is not possible to transform to a frame where both detectors are simultaneously at rest. In this paper, we address this issue, and analyze how the relative velocity between a pair of UDW detectors affects their ability to harvest entanglement. By introducing an additional variable, we obtain an additional degree of freedom which improves the ability of the detectors to harvest entanglement under the right circumstances. 
Experimentally speaking, the
velocity of a detector  is much easier to control than 
its acceleration, thus making it a versatile degree of freedom.

 We begin in section \ref{sec:setup} with the set-up of the situation and the general formalism for the point-like two level UDW detector evolving in vacuum. The novel results of this paper are presented in section \ref{results}, and we end with conclusions in section \ref{conclusion}.

All work are done in natural units $\hbar = G=c=1$ and flat space-time. 

\section{Set-up}
\label{sec:setup}

The UDW detector is a two-level qubit with an energy gap $\Omega$ between its
ground and excited states that couples linearly to a scalar field $\phi$.
%analogous to that of a hydrogen atom.
Despite its simplicity, this model captures the essential features of light-matter interaction when no angular momentum is exchanged \cite{Lopp:2020qwx}. It also allows us to straightforwardly and quantitatively model the harvesting of entanglement \cite{Valentini1991nonlocalcorr,reznik2003entanglement, Steeg2009,Salton:2014jaa} from the field.
 
To study the dependence of entanglement harvesting on the relative velocity between two detectors, we use a set-up consisting of two UDW detectors with identical $\Omega$, each moving at some constant velocity in Minkowski spacetime. 
The system is initialized with both detectors in their ground states and the scalar field in the vacuum state.
Each detector has a switching function (see below), which describes the duration and strength of the interaction between each detector and the field as a function of time. We will use a Gaussian for the switching function. The detectors  evolve according to the UDW/scalar Hamiltonian, and after some time,   become entangled due to their interactions with the quantum scalar field. We are interested in measuring the amount of entanglement between the detectors as a function of the parameters of the detectors and their relative velocity.

The two detectors will henceforth be labeled using $D\in\{A,B\}$; the Hilbert space of each detector is $\mathcal{H}_D = \text{span}\{\ket{0}_D,\ket{1}_D\}$ where $\ket{0}_D$ and $\ket{1}_D$ denote their ground and excited states,  separated by an energy gap $\Omega_D$. The Hilbert space of the scalar quantum field is denoted by $\mathcal{H}_\phi$, with Fock eigenstates $\{\ket{n}:n\in \mathbb{N}\}$.

Consider a detector travelling in Minkowski spacetime with the trajectory $\textbf{x}_D(t) = (t,\Vec{x}(t))$, with $\Vec{x}=(x_1,x_2,x_3)$, parametrized by some observer's time 
$t$. The interaction Hamiltonian between  detector $D$ and the scalar field $\hat{\phi}[\textbf{x}]$ is given by \cite{PhysRevD.14.870, DeWitt:1980hx}
\begin{equation}
    H_D (\tau_D)=\lambda  \chi(\tau_D)\mu_D(\tau_D)\otimes\hat{\phi}[\textbf{x}_D(t)]
\end{equation}
where $\tau_D$ is the proper time of  detector $D$, $\chi(\tau_D)$ its  switching function, and $\lambda$ is the detector-field coupling constant. The quantity $\mu_D=(e^{i\Omega_D \tau_D}\sigma_D^++e^{-i\Omega_D \tau_D}\sigma_D^-)$ is the monopole moment of   detector $D$, with $\sigma^\pm_D$ being the respective raising and lowering operators.

Using the frame of reference of some inertial observer, a complete Hamiltonian of the detector pair and scalar field can be obtained. We set the observer's time $t=0$ to coincide with $\tau_D=0$, which gives the relation $\tau_D(t)=t/\gamma_D$, where $\gamma_D$ is the Lorentz factor between detector and observer. As the detectors do not directly interact with each other, we can write the full interaction Hamiltonian as
\begin{equation}
    H_I(t)=\frac{d\tau_A}{dt}H_A(\tau_A)+\frac{d\tau_B}{dt}H_B(\tau_B) \,.
\end{equation}
The initial state of the detector-field system is set to be $\ket{\psi}_0:=\ket{0}_A\ket{0}_B\ket{0}_\phi$, with $\ket{0}_\phi$   the vacuum state of the field.

The initial state time-evolves to $\ket{\psi}_f=U\ket{\psi}_0$, where
 \begin{equation}
    U:=\mathcal{T}\bigg[\textbf{Exp}\bigg(-i\int_{-\infty}^\infty dt \sum_{D\in\{A,B\}}\frac{d\tau_D}{dt}H_D(\tau_D)\bigg)\bigg]\,,
\end{equation}
with $\mathcal{T}$ being the time ordering operator. We are only interested in the final state of the detector pair, which is described by the density matrix $\rho_{AB}\in \mathcal{S}(\mathcal{H}_A\otimes \mathcal{H}_B)$, obtained by taking the partial trace over the field of the final state, i.e., $\rho_{AB}= \tr_{\phi}(\ket{\psi}_f\bra{\psi}_f)$.
Computing all quantities to second order in $\lambda$, we obtain
\cite{Pozas2015}
\begin{equation}
    \rho_{AB}=
    \begin{pmatrix}
1-P_A-P_B & 0 & 0 & X\\
0 & P_A & C & 0\\
0 & C^* & P_B & 0\\
X^* & 0 & 0 & 0
\end{pmatrix}+\mathcal{O}(\lambda^4)
\label{densm}
\end{equation}
where
\begin{widetext}
\begin{align}
    P_D&:=\lambda^2\int_{-\infty}^\infty d\tau_D d\tau'_D \chi_D(\tau_D)\chi_D(\tau'_D)W(\textbf{x}_D(t), \textbf{x}_D(t'))e^{-i\Omega_D (\tau_D-\tau'_D)} 
    \label{2.6}
    \\
    C&:=\lambda^2\int d\tau_A(t)d\tau_B(t')\chi_A(\tau_A(t))\chi_B(\tau_B(t'))e^{-i(\Omega_A\tau_A(t)-\Omega_B\tau_B(t'))}W(\textbf{x}_A(t),\textbf{x}_B(t'))
    \\
    X&:=-\lambda^2\int_{t>t'} d\tau_A(t) d\tau_B(t')\bigg[\chi_A(\tau_A(t))\chi_B(\tau_B(t'))e^{-i(\Omega_A\tau_A(t)+\Omega_B\tau_B(t'))}W(\textbf{x}_A(t'),\textbf{x}_B(t))\nonumber 
    \\
    &\qquad +\chi_A(\tau_A(t'))\chi_B(\tau_B(t))e^{-i(\Omega_A\tau_A(t')+\Omega_B\tau_B(t))}W(\textbf{x}_B(t'),\textbf{x}_A(t))\bigg]\,.
\end{align}
\end{widetext}
Here $W(\textbf{x},\textbf{x}')=\bra{0}\phi(\textbf{x})\phi(\textbf{x}')\ket{0}$ is the two-point correlator or the \textit{Wightman function} of the field. If the density matrix in \eqref{densm} is further reduced to that of one of the detectors (obtained by tracing over the other), we obtain the expression $\rho_D=(1-P_D)\ket{0}\bra{0}_D+P_D\ket{1}\bra{1}_D$, indicating that $P_D$ is the transition probability of detector $D$.
The quantities $C$ and
$X$ describe correlations between   between $A$ and $B$, with the
latter containing non-local correlations associated with   entanglement, as we shall see.

To quantify the entanglement between the detectors, we can use
\begin{equation}
    \mathcal{N}=\textbf{max}\bigg(\sqrt{|X|^2+\bigg(\frac{P_A-P_B}{2}\bigg)^2}-\frac{P_A+P_B}{2},0\bigg) 
    \label{Neg}
\end{equation}
which is the
\textit{Negativity}, an entanglement monotone that is equivalent to the concurrence in the particular case of qubit detectors  with identical gaps \cite{Mart_n_Mart_nez_2016}. 
 
We will use for the switching functions Gaussian distributions, with widths $\sigma_A=\sigma_B =\sigma$ that each peaks at $\tau_D=0$ 
\begin{equation}
    \chi(\tau_D)=e^{-\tau_D^2/4\sigma^2}\,
    \label{chi}
\end{equation}
in the rest frame of the detector. Apart from being experimentally realizable (at least in principle \cite{Stritzelberger_2021})  
this kind of switching function is convenient insofar as  it is analytic and allows for the far end of the distribution to be cut off with relatively small consequences \cite{Mart_n_Mart_nez_2016}. 

As the entanglement is a Lorentz-invariant quantity, and
we are dealing with inertial detectors, we can work in the center-of-mass frame without loss of generality.    In this frame, the two detectors   move at opposite velocities of equal magnitude $v$, and their proper times are equal:
$\tau_A=\tau_B=\tau=t/\gamma= t\sqrt{1-v^2}$.  We set $t=0$ to coincide with the moment of closest approach of the detectors, which is when the switching functions are maximized.
We arrange the detectors so that they are offset in a direction orthogonal to their direction of motion; at $t=0$ the displacement between them is minimal and denoted by $d$. 

In this paper, we study the scenario in which $\Omega_A=\Omega_B=\Omega$.
Consequently,  we can get rid of the subscript on the probability and denote it simply as $P\equiv P_A=P_B$. By applying these conditions to ~\eqref{chi}, ~\eqref{prob} and~\eqref{Neg} we obtain
\begin{equation}
    \mathcal{N}=\textbf{max}\bigg(|X|-P,0\bigg)+\mathcal{O}(\lambda^3)\,,
    \label{maxn}
\end{equation}
with
\begin{align}
P&=\lambda^2\int_{-\infty}^\infty d\tau e^{-i\Omega \tau}e^{-\tau^2/2\sigma^2} \int_{-\infty}^\infty d\tau' e^{i\Omega \tau'}e^{-\tau'^2/2\sigma^2}
\nonumber
\\
&\quad \times W(\textbf{x}_D(\gamma \tau), \textbf{x}_D(\gamma\tau'))
\label{probt}
\\
    X &=-\frac{1}{2}\lambda^2\int_{-\infty}^\infty du \,e^{-u^2/4\sigma^2} e^{-i\Omega u}\int_0^\infty ds\,e^{-s^2/4\sigma^2}
    \nonumber
    \\
    &\quad \times\bigg[W(\textbf{x}_A(\gamma(u-s)/2),\textbf{x}_B(\gamma(u+s)/2)) \nonumber
    \\
    & \qquad + W(\textbf{x}_B(\gamma(u-s)/2),\textbf{x}_A(\gamma(u+s)/2))\bigg]
    \label{corrt}
\end{align}
where we have opted to substitute $t_d$ with $\gamma \tau_d$ and $u=\tau_A(t)-\tau_B(t')$ and $s=\tau_A(t)-\tau_B(t')$.

%\section{3+1 Scalar Coupling}\label{4D}

In $(3+1)D$ Minkowski space  the scalar field and Wightman function may be written as \cite{Birrell:1982ix}
\begin{align}
    \phi(t,\Vec{x})&=\int{\frac{dk^3}{(2\pi)^{3/2}}\frac{1}{\sqrt{2|\Vec{k}|}}\bigg(e^{-i|\Vec{k}|t+i\Vec{k}\cdot \Vec{x}}a_{\Vec{k}}+e^{i|\Vec{k}|t-i\Vec{k}\cdot \Vec{x}}a_{\Vec{k}}^\dag}\bigg)\\
W(\bold{x},\bold{x'})&=-\frac{1}{4\pi^2}\lim_{\epsilon\to 0^+}\frac{1}{[t-t'-i\epsilon\,\textbf{sgn}(t-t')]^2-|\Vec{x}-\Vec{x}'|^2} 
\label{wightman3}
\end{align}
with the $\epsilon$ term in~\eqref{wightman3}   a UV-regulator.

\begin{figure}
    \centering
    \includegraphics[width=8cm]{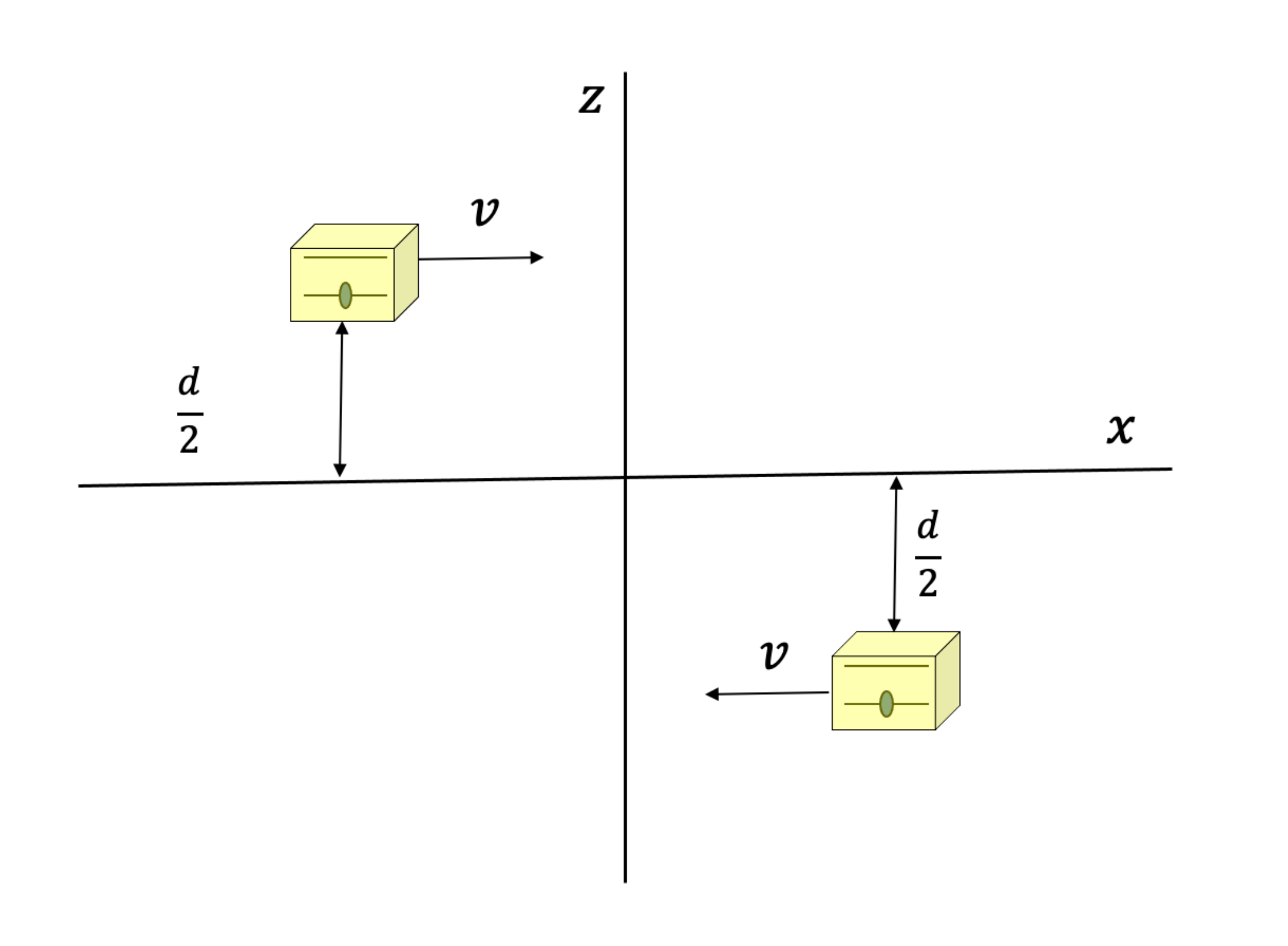}
    \caption{Basic setup of the two detectors, which begin their trajectories at arbitrarily large values of $|x|$.
 }
    \label{setup}
\end{figure}

Using the center-of-mass reference frame, we set the origin and orientation such that detectors $A$ and $B$ are traveling respectively at speeds $v,-v$ in the $x_1$ direction, with distance of closest approach $d$ occurring at $t=0$. This yields the trajectories
\begin{equation}\label{trajs}
     \Vec{x}_A(t)=\bigg(vt,0,\frac{d}{2}\bigg), \quad
     \Vec{x}_B(t)=\bigg(-vt,0,-\frac{d}{2}\bigg)=-\Vec{x}_A(t) 
 \end{equation}
 depicted in fig. \ref{setup}.
  The Wightman function \eqref{wightman3} when evaluated along the trajectories of $A$ and $B$ gives
  \begin{widetext}
\begin{equation}\label{Wightman}
    W(\bold{x_A}(t'),\bold{x_B}(t))=-\frac{1}{4\pi^2}\lim_{\epsilon\to 0^+}\frac{1}{[t-t'-i\epsilon\,\textbf{sgn}(t-t')]^2-(d^2+v^2(t^2+t'^2))}\,,
\end{equation}
\end{widetext}
yielding the expressions
\begin{align}
    P&=\frac{\lambda^2}{4\pi}[e^{-\sigma^2\Omega^2}-\sqrt{\pi}\sigma\Omega\; \textbf{erfc}(\sigma\Omega)] \,,
    \label{prob}
    \\
X&=\lambda ^2\left(\frac{1-v^2}{8\pi i}\right)\int _{-\infty }^{\infty }du\,\,
   \frac{\text{Exp}\left[-\frac{d^2(1-v^2)+u^2(1-v^4)}{4\sigma
   ^2}\right]}{\sqrt{v^2u^2+d^2}} 
   \nonumber
   \\
   &\quad \times e^{-i\Omega u\sqrt{1-v^2}}\left(1+\textbf{erf}\left[\frac{i    
   \sqrt{1-v^2}\sqrt{v^2u^2+d^2}}{2\sigma }\right]\right)
   \label{corr}
\end{align}
upon substitution   into~\eqref{probt} and \eqref{corrt}.   We shall sometimes
find it convenient to write $\mathcal{N}=\max\{\mathcal{M},0\}$ where
$\mathcal{M}\equiv|X|-P$.

  We note that our Gaussian switching function has infinite support, which means that detector $A$ can always send a light-signal at a time $\tau_A$ when $\chi(\tau_A)\neq 0$ which will be received by $B$ at a time $\tau_B$ when $\chi(\tau_B)\neq 0$. In other words the detectors are, strictly speaking, always within causal contact. However, since the Gaussian function is exponentially suppressed at times away from the peak, it approximates a compact switching function supported within 
  $\pm 3\sigma$ around its peak. 
  We shall regard the detectors to be ``spacelike'' separated when a signal sent within $\pm 3\sigma$ of detector $A$'s switching peak cannot be received by $B$ within $\pm 3\sigma$ of $B$'s switching peak. Clearly in the case of static detectors separated by a coordinate distance of $d$ in Minkowski, the detectors are spacelike separated if $d>6\sigma$. 
  
In the present case, using \eqref{trajs},
  for a vector $\vec{r} = x_B(t_2)-x_A(t_1)$ (the tangent vector to a null geodesic joining a point on $A$ 
  at time $t_1$ 
  to a point on $B$ at time $t_2$) to be light-like we require $-(t_2-t_1)^2 = v^2(t_2+t_1)^2 + d^2$. 
  This gives
$$t_2 = \frac{(1+v^2)t_1+\sqrt{d^2(1-v^2)+4t_1^2v^2}}{1-v^2}\,,$$that is, a light signal sent at $t=t_1$ from $A$ will reach $B$ at the above $t_2$. Observe that $t_2$ increases as $t_1$ increases, so $A$ can only send signals to $B$ if a signal sent at  (accounting for proper time) $t_1=-3\sigma/ {\sqrt{1-v^2}}$ can reach $B$ at $t_2=3\sigma/ {\sqrt{1-v^2}}$. This translates into the criteria:
\begin{widetext}
$$\frac{3\sigma}{{\sqrt{1-v^2}}} > \frac{(1+v^2)(-\frac{3\sigma}{ {\sqrt{1-v^2}}})+\sqrt{d^2(1-v^2)+4\left(-\frac{3\sigma}{ {\sqrt{1-v^2}}}\right)^2v^2}}{1-v^2}\,,$$
\end{widetext}
which has the solution $d\leq \frac{6\sigma}{{\sqrt{1-v^2}}}$. Hence the criteria for spacelike separation is 
 $d\geq\frac{6\sigma}{\sqrt{1-v^2}}$.

\section{Results and Discussion}  
\label{results}

\subsection{Static Case}

The behavior of the Negativity at constant detector separation and $v=0$, $\mathcal{N}(v=0)$, has been well studied \cite{Pozas_Kerstjens_2016}. Here we briefly review these properties, writing
$\mathcal{N}(v)$ to denote the dependence of the Negativity on $v$.
In particular, we have
\begin{align}
    \mathcal{N}(0)&=\max\bigg\{\underbrace{\frac{\lambda^2\sigma}{4d\sqrt{\pi}}e^{-\frac{d^2}{4\sigma^2}}e^{-\sigma^2\Omega^2}\sqrt{1+\textbf{Erfi}\left[\frac{d}{2\sigma}\right]^2}}_{|X|}
    \nonumber
    \\
    &
    \quad 
    -\underbrace{\frac{\lambda^2}{4\pi}[e^{-\sigma^2\Omega^2}-\sqrt{\pi}\sigma\Omega\textbf{Erfc}(\sigma\Omega)]}_{P},0\bigg\}
    \label{zeron}
\end{align}
\begin{figure}
    \centering
    \includegraphics[width=0.9\linewidth]{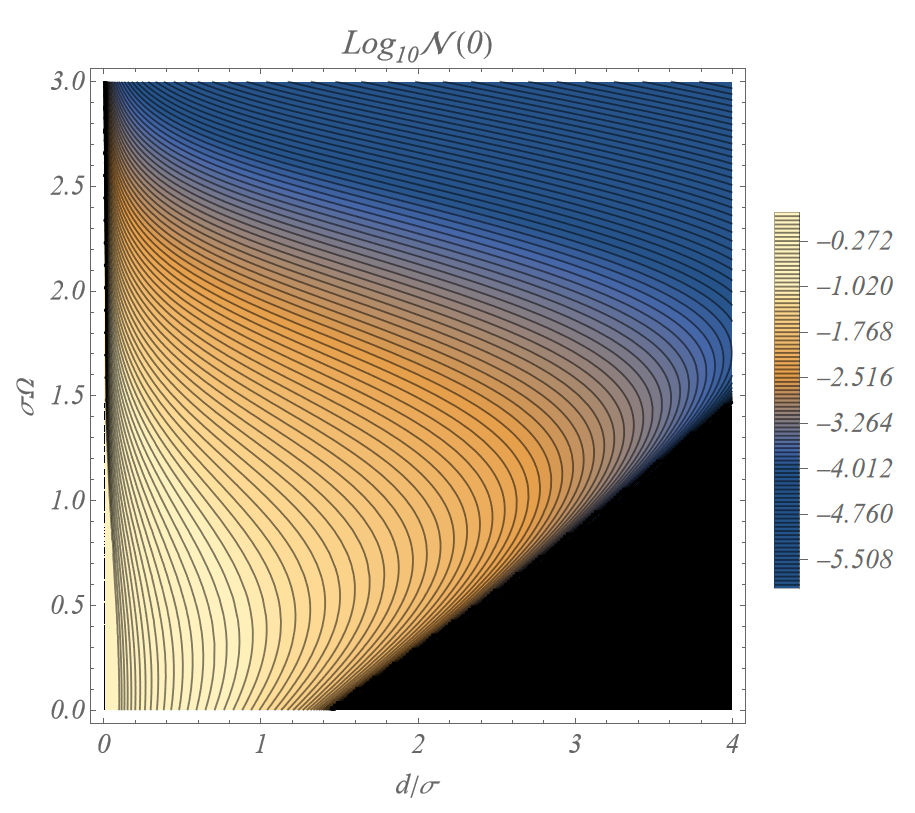}
    \caption{The logarithmic contour plot of $\mathcal{N}(0)$ in units of $\lambda^2$ as function of $d/\sigma$ and $\sigma\Omega$. The black region on the bottom right triangle represents a region of zero entanglement in the parameter space.
 }
    \label{fig:my_label}
\end{figure}
It can be easily shown that $\mathcal{N}(0)$ in \eqref{zeron} is a strictly decreasing function of distance, asymptotically approaching $\infty$ as $d/\sigma\to 0$,  
and 0 as $d/\sigma\to \infty$. This is illustrated in fig. \ref{fig:my_label}. Note that for  $d/\sigma \lesssim \lambda^{2}$ our approximation becomes flawed, as   higher order terms $\mathcal{O}(\lambda^4)$ become relevant. Strictly speaking we can't arbitrarily decrease distance to increase entanglement. The derivative of
$\mathcal{M}$  with respect to $\Omega$ is strictly positive at $\sigma\Omega=0$. Hence around $\sigma\Omega=0$, unless $\mathcal{N}(0)$ is locally zero (i.e $\mathcal{M}<0$), $\mathcal{N}(0)$ increases correspondingly with $\sigma\Omega$.

\subsection{Moving Case}

We now analyse the effects of varying $v$ and $d$ on the entanglement harvested between the two detectors. Negativity will be calculated in units of $\lambda^2$ while the parametric variables ${d,\Omega}$ are scaled into the unitless quantities ${d/\sigma,\sigma\Omega}$.

\begin{figure} 
    \centering       
        \includegraphics[width=8cm]{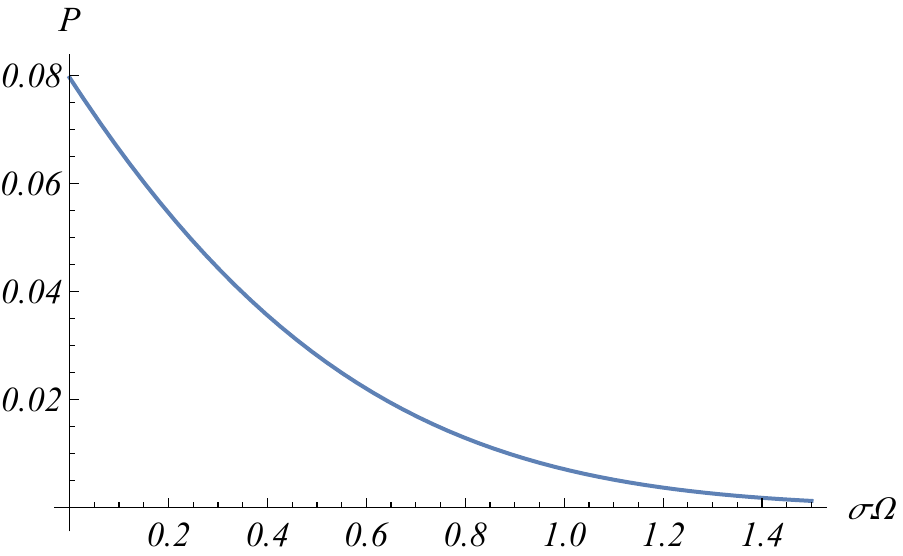}
\caption{Plot of $P$ in units of $\lambda^2$ against $\sigma\Omega$.}
\label{fig:probgraph}
\end{figure}
 As seen in fig. \ref{fig:probgraph}, the excitation probability $P$ monotonically decreases as $\sigma \Omega$ increases, asymptotically approaching zero, while remaining positive throughout. This profile is identical to the static case, and in particular, it is independent of the relative velocity $v$ and is indistinguishable from the profile of static detectors, as expected from Lorentz invariance. 
  
\begin{figure*} 
\label{NvsV}
    \includegraphics[width=0.4\linewidth]{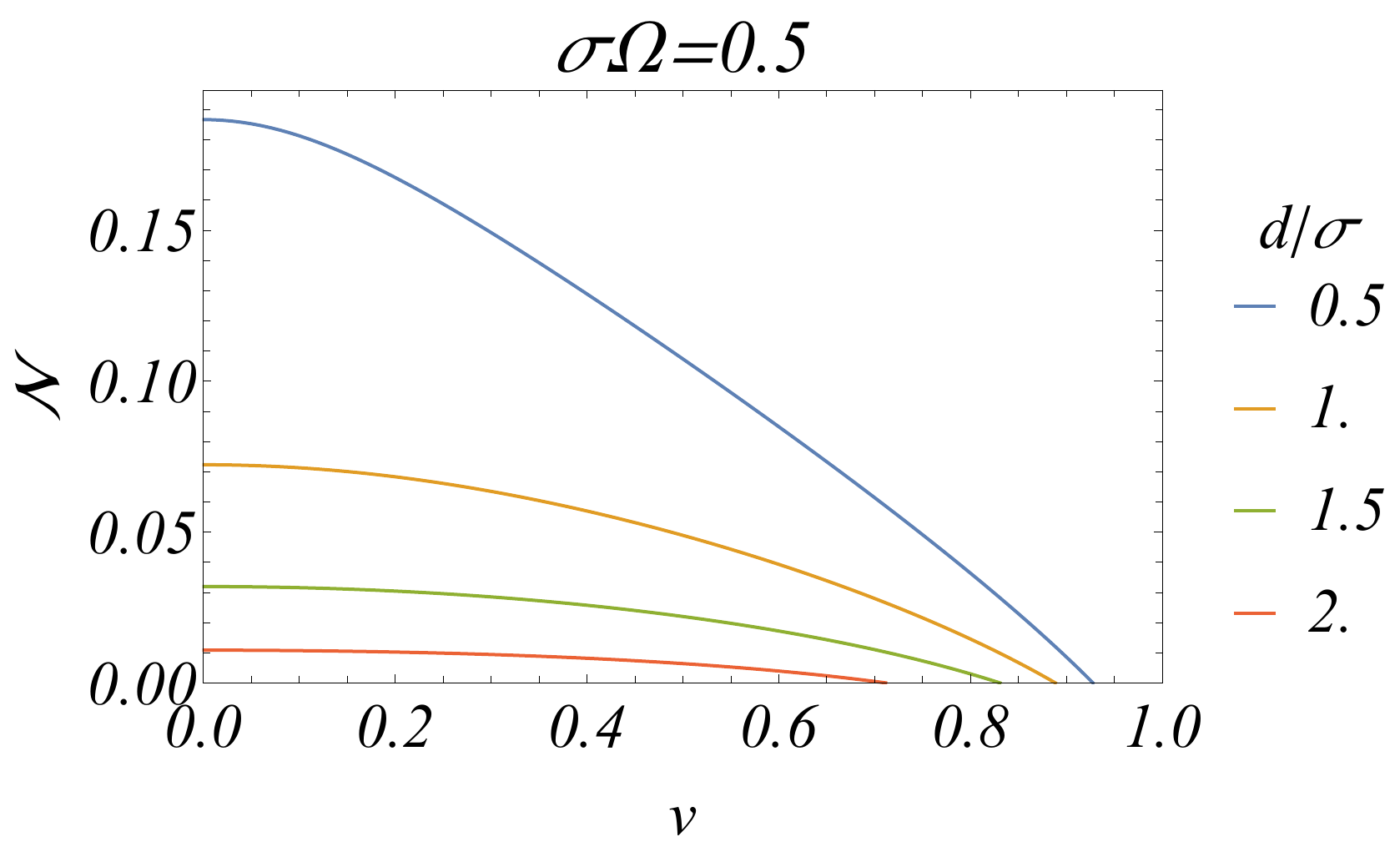} 
    \includegraphics[width=0.4\linewidth]{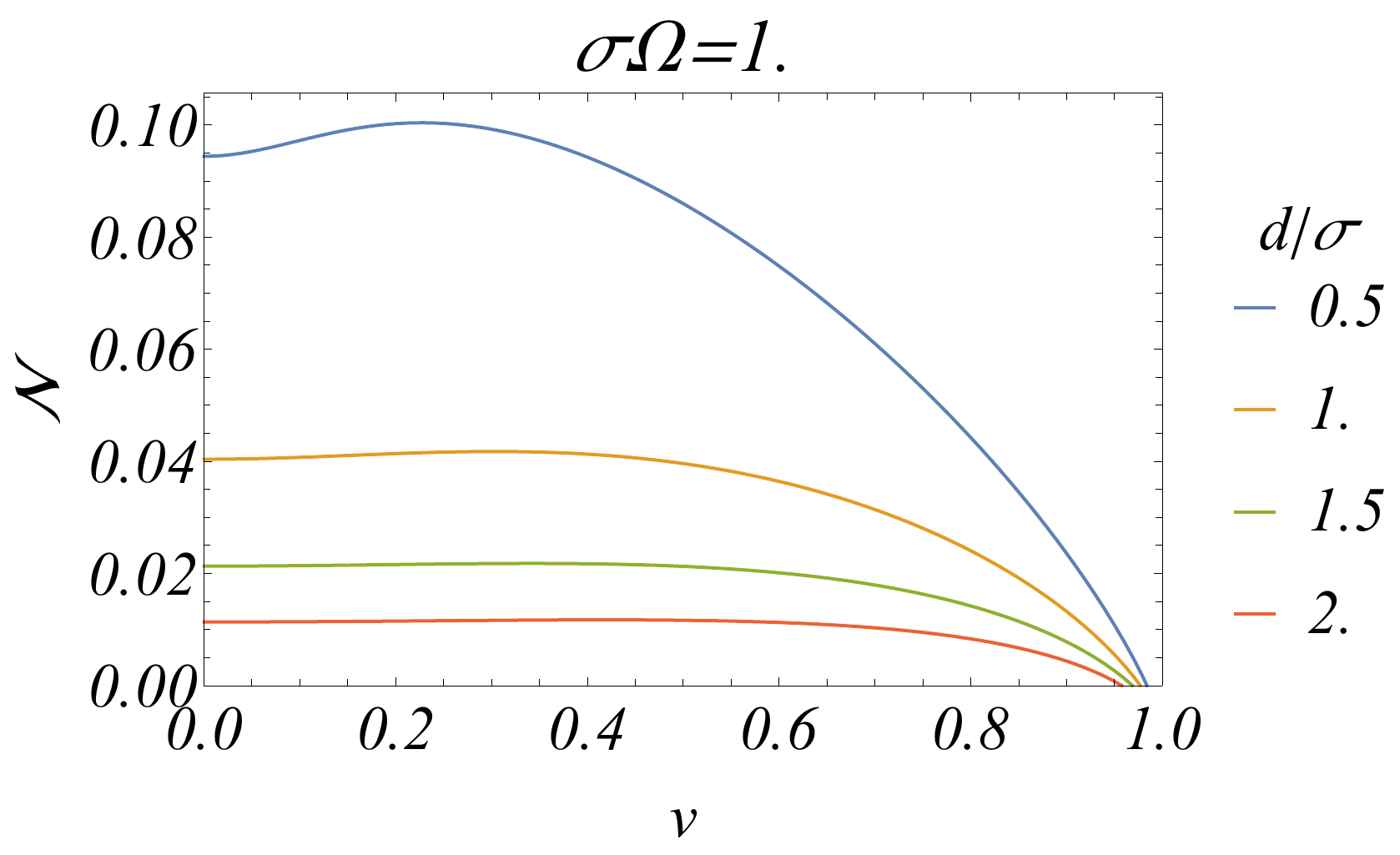} 
    \includegraphics[width=0.4\linewidth]{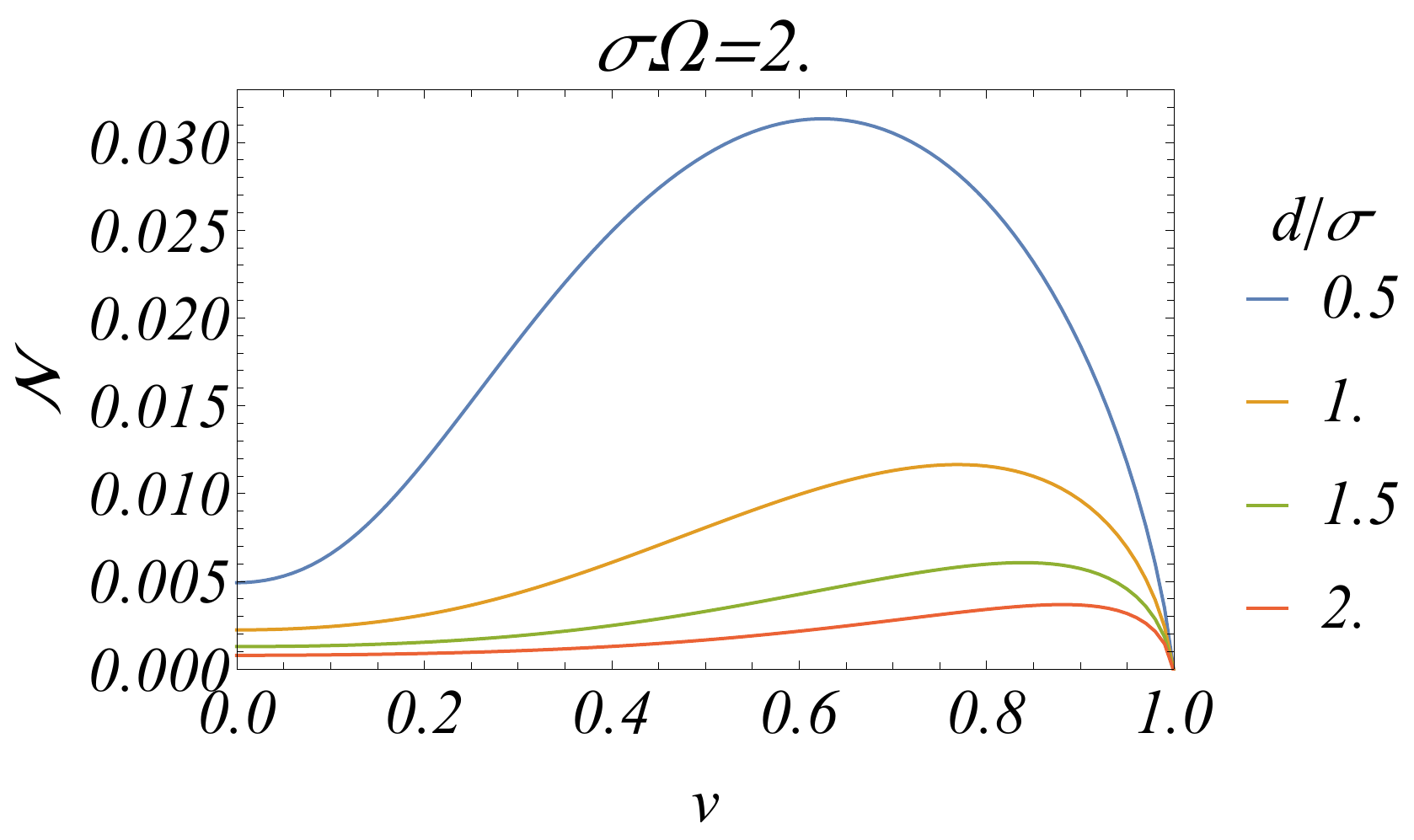} 
    \includegraphics[width=0.4\linewidth]{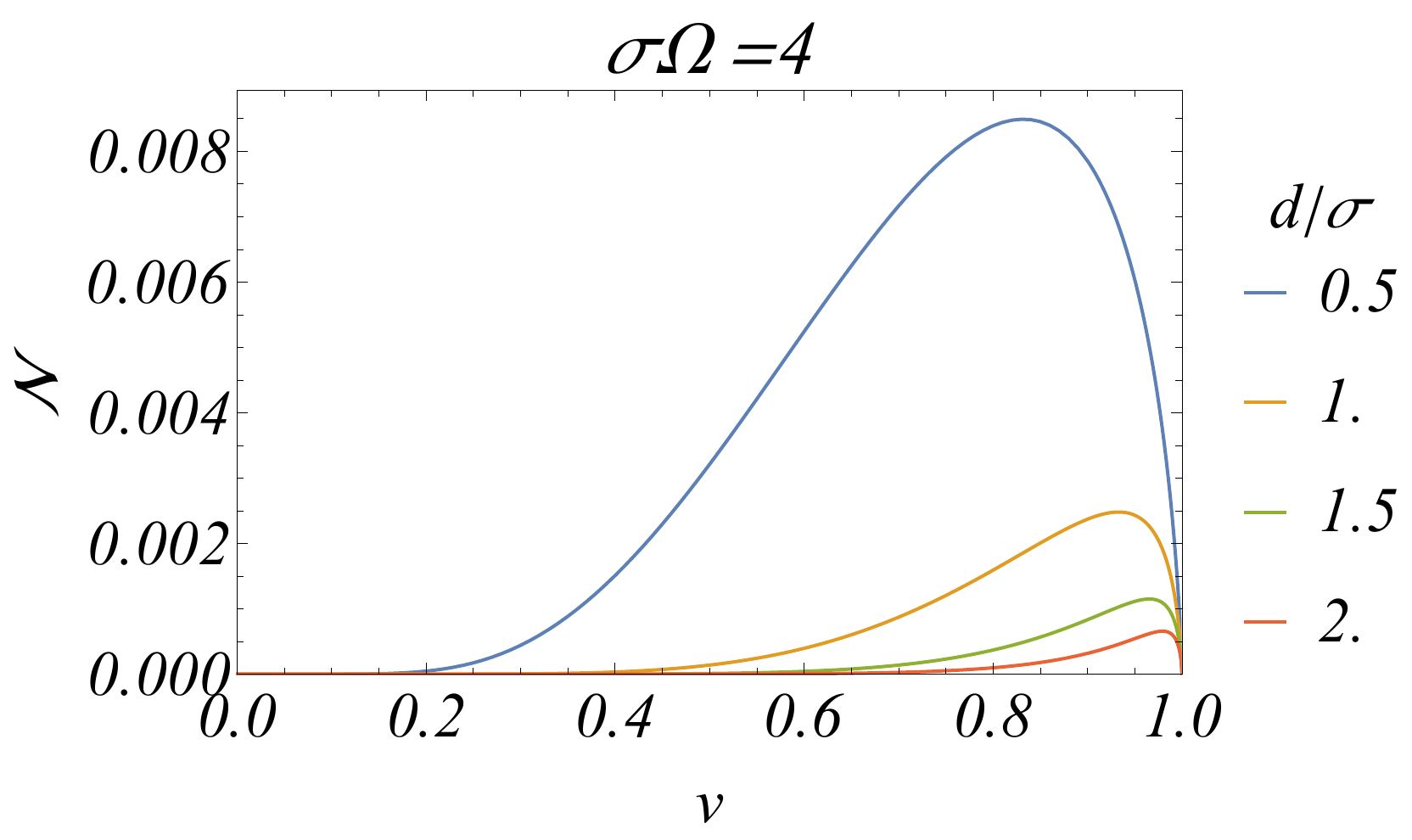} 
 \caption{Plot of Negativity $\mathcal{N}$ in units of $\lambda^2$ as a function of velocity $v$ with varying minimal distance $d/\sigma$, at $\sigma\Omega=0.5,1,2,4$. }
 \label{fig:lineplot}
\end{figure*}

  In fig. \ref{fig:lineplot} we plot
  Negativity as a function of $v$ for various separations.  We observe the expected result that entanglement always decreases with $d/\sigma$ at fixed $v$. However we observe the surprising result that Negativity is maximized for a particular gap-dependent choice of $v$. The effects are particularly dramatic for small $d$  as the gap increases. Entanglement does not correspondingly always decrease as energy gap increases, consistent with the static case (fig. \ref{fig:my_label}). Finally, we observe from fig. \ref{fig:lineplot}
  that entanglement is extinguished at sufficiently large values of $v$ for all $d$ and $\Omega$.  As $d$ increases this takes place at progressively smaller velocities (evident in the upper left panel in fig. \ref{fig:lineplot}), but as $\Omega$ increases the values of $v$ at which $\mathcal{N}\to 0$ and the positions of the peaks (provided they exist) cluster closer and closer to $v=1$.  
  
\begin{figure*}
\includegraphics[width=0.45\linewidth]{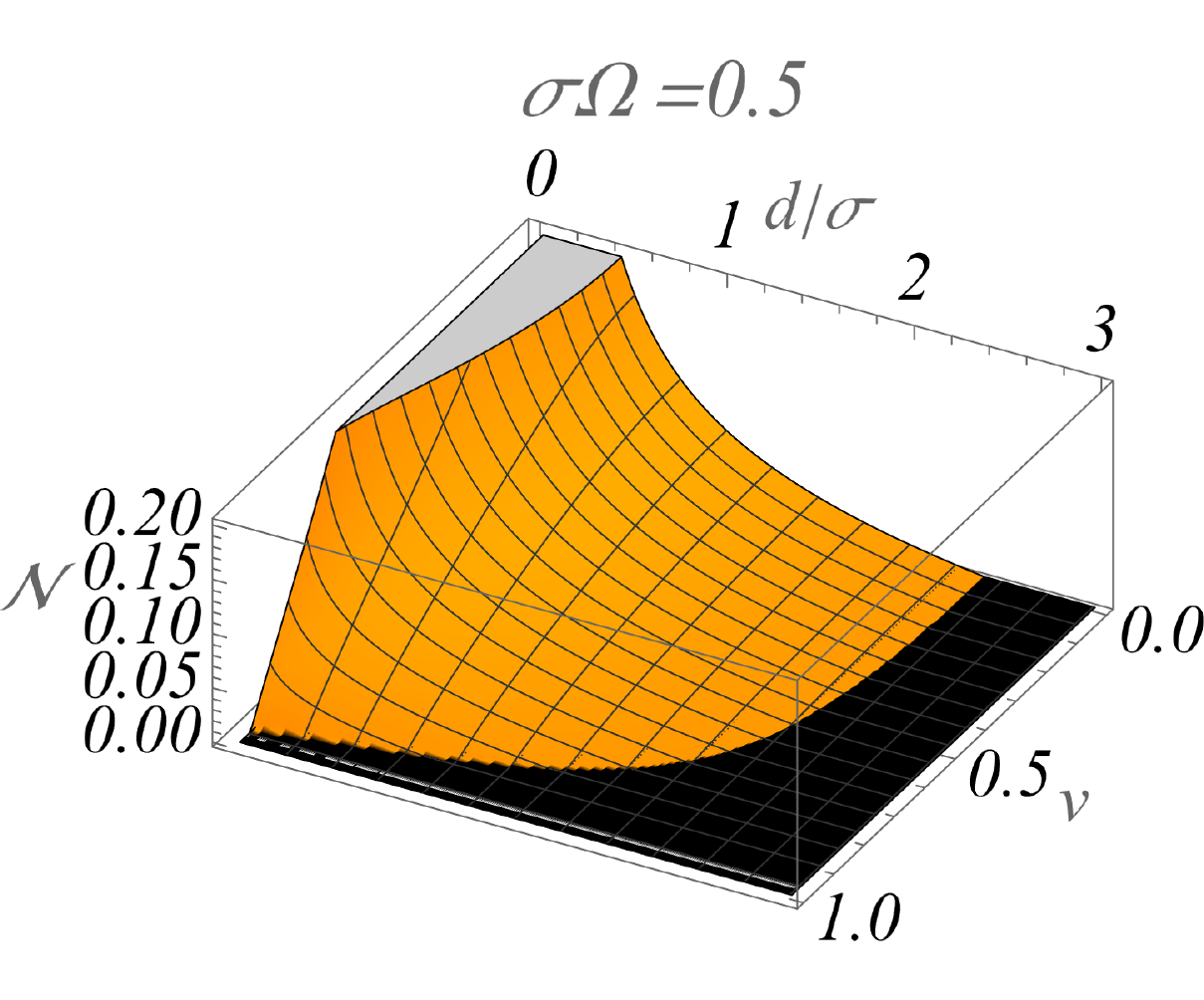} 
    \includegraphics[width=0.45\linewidth]{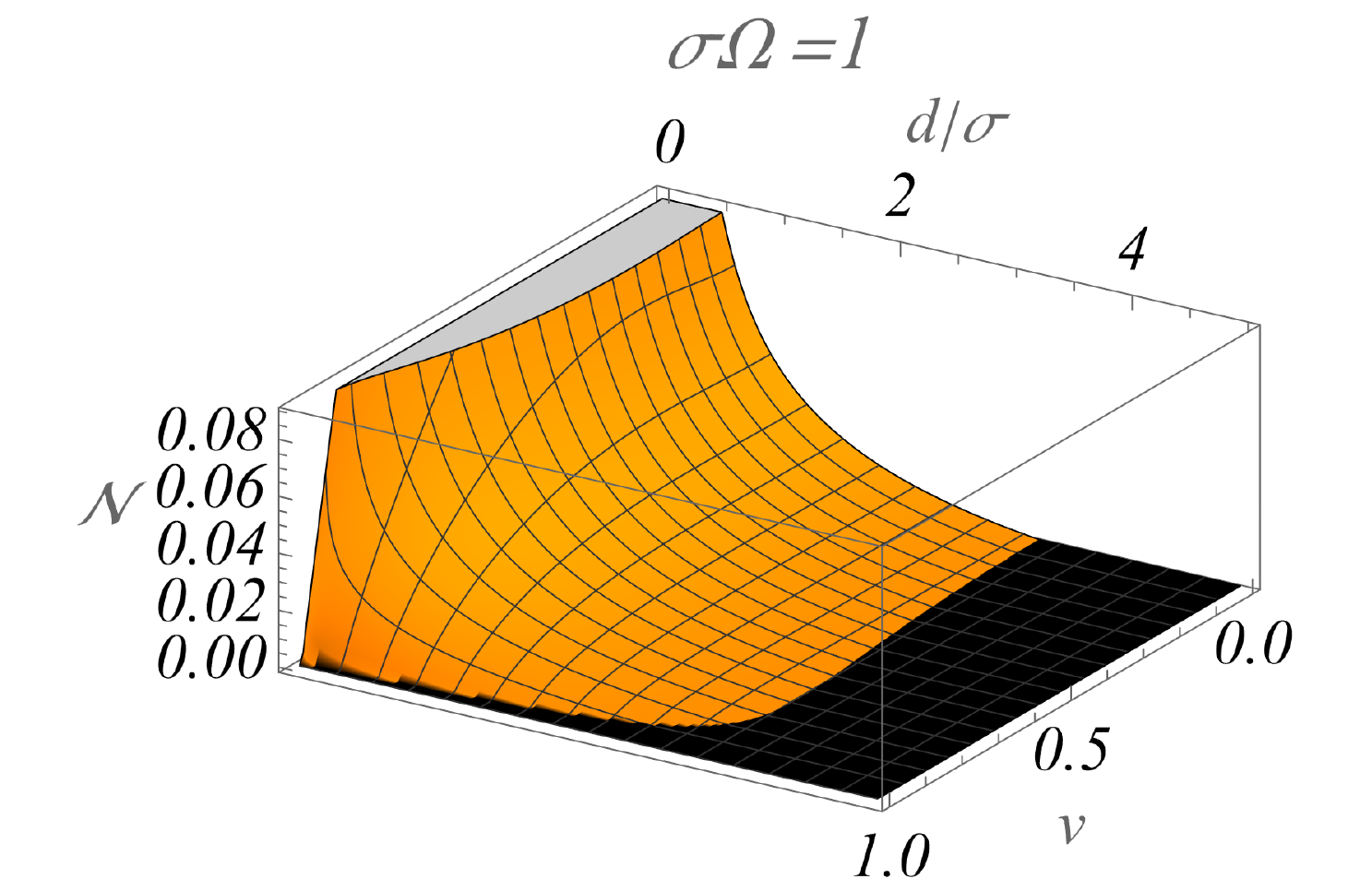} 
    \includegraphics[width=0.45\linewidth]{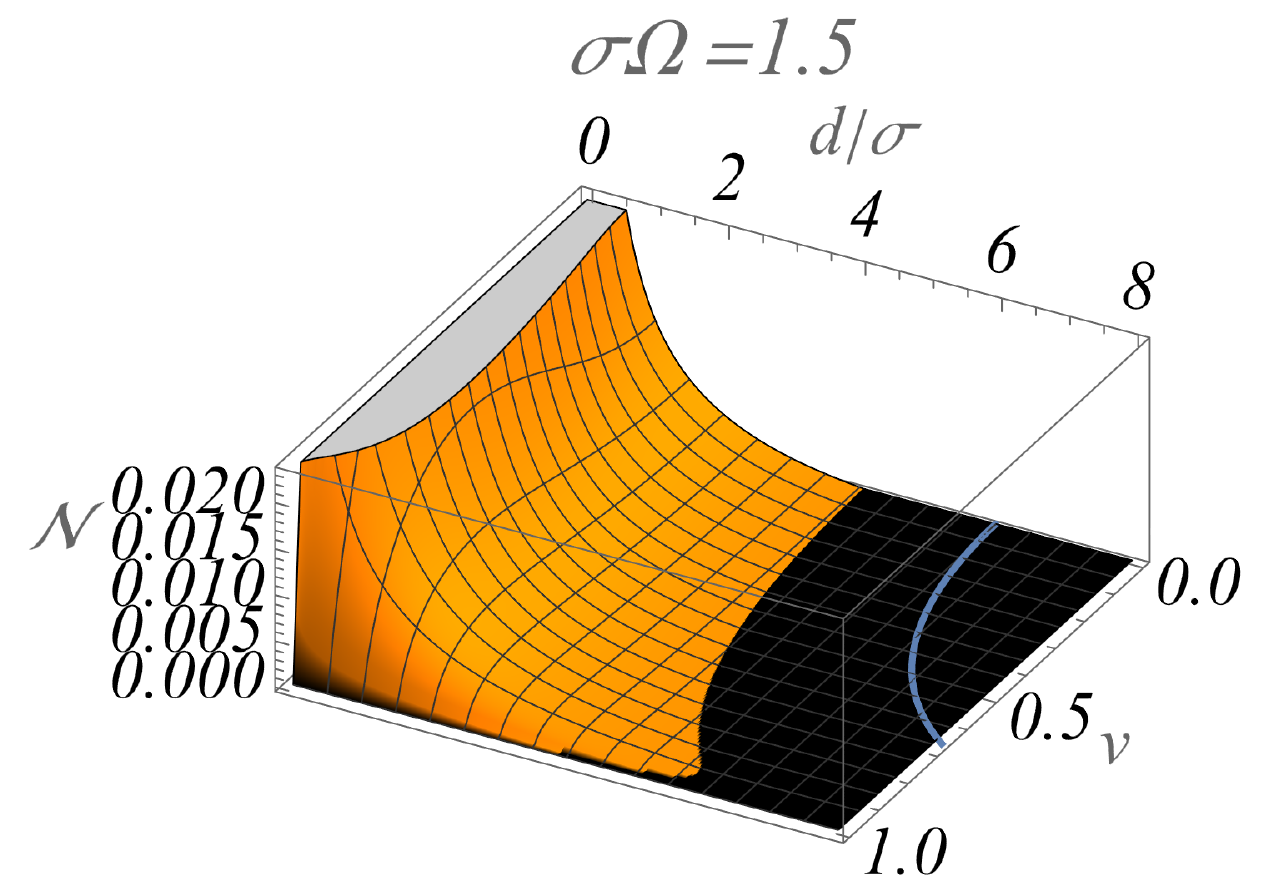}
    \includegraphics[width=0.45\linewidth]{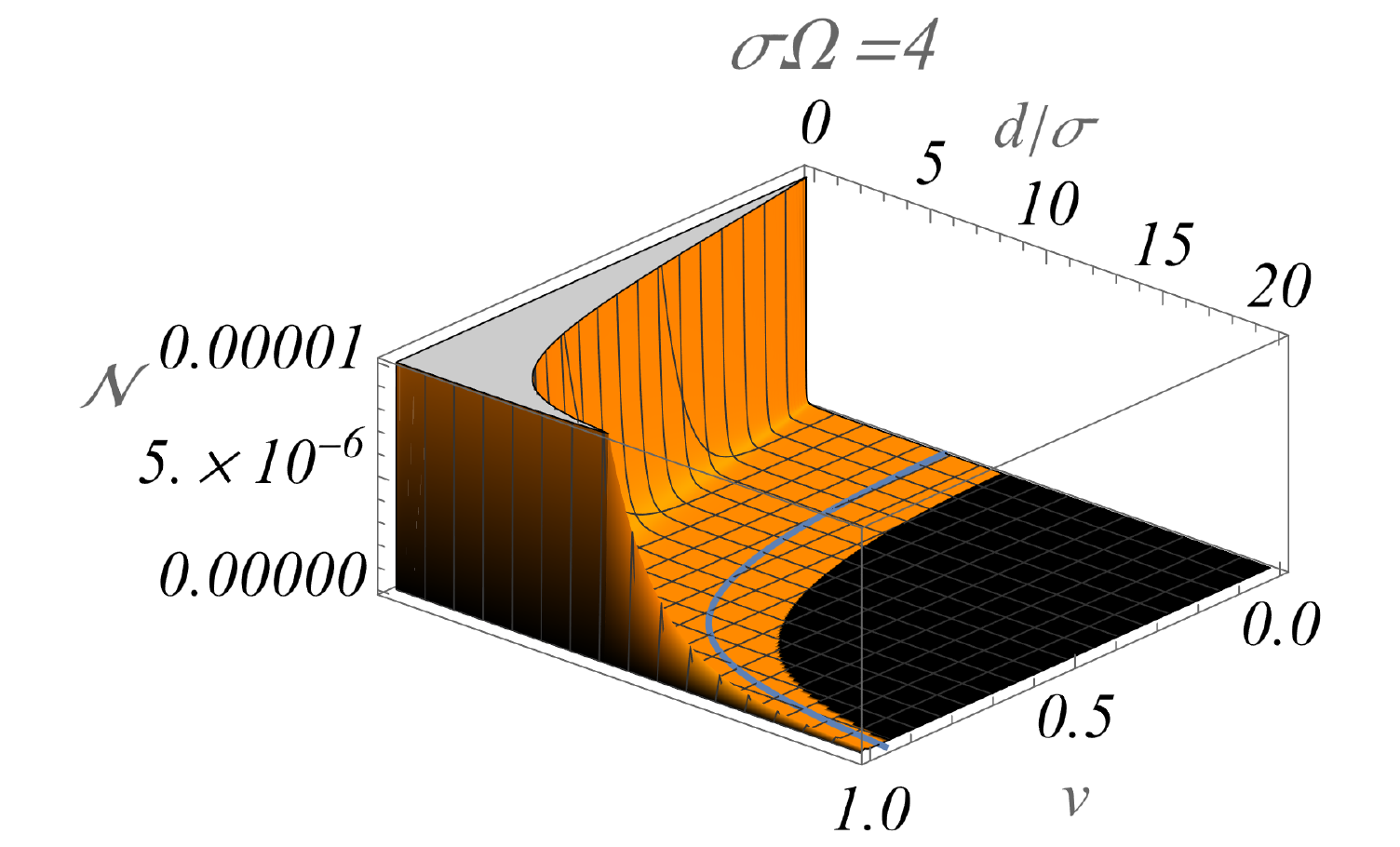} 
 \caption{$3D$ plots of Negativity in units of $\lambda^2$ at $\sigma\Omega=0.5,1,1.5,4$ as a function of minimal distance $d$ and velocity $v$. The light cone criterion $d = \frac{6\sigma}{\sqrt{1-v^2}}$ -- the maximum distance within which the detectors are considered to be timelike separated -- is denoted by the  blue line. We see that the positive entanglement zone in the $\sigma\Omega=0.5,1$ cases are all within the light cone; note that this
 boundary is outside the ranges plotted in  the upper two graphs. Note that spacelike entanglement
 becomes evident for
 $\sigma\Omega=4$. As the cross sectional plot of fig. \ref{fig:lineplot} shows, the negativity in spikes considerably near $v=1$ and proceeds to continuously drop to zero at some high velocity.}
 \label{fig:3Dplot}  
\end{figure*}

 The $\sigma\Omega=4$ case in fig. \ref{fig:lineplot} and fig. \ref{fig:3Dplot} reveals that entanglement can be harvested outside of our lightcone criterion. The ability of detectors to harvest spacelike entanglement is known in the literature and has been studied for example in the static case in \cite{Pozas_Kerstjens_2016}
 and with indefinite temporal order
\cite{Henderson:2020zax}. At any fixed energy gap,  we note that it is always impossible to harvest entanglement given a sufficiently high minimal distance. 

  The dependence of entanglement on $v$ can be understood as follows: the correlation factor $X$ asymptotically approaches zero as $v$ increases, while the probability remains at a fixed positive number. Hence at arbitrary high speed, the amount of noise due to spontaneous excitation of the detectors will outweigh the correlations. 
  We plot the joint effects of
  $d$ and $v$ on Negativity in fig. \ref{fig:3Dplot} for various values of $\Omega$. We see the region of vanishing Negativity moving closer to $v=1$ as $\Omega$ increases. Conversely, this region moves closer to $v=0$ as $d$ increases,   evident in fig. \ref{fig:lineplot}. The vanishing of entanglement in the limit $v\rightarrow 1$ agrees with results in \cite{Koga:2018the,Barman:2022xht}.
  
It might come as a surprise that we can increase entanglement by increasing $v$. The static case shows that Negativity is inversely proportional to the distance between the detectors. Intuitively, a higher velocity   increases the average distance between the two detectors during their interaction with the field,  resulting in lower entanglement. However this does not take into consideration the effect  of time dilation, which enhances the non-local correlations \eqref{corr} for a  region in the $(v,\Omega)$ parameter space. More analysis of this will be given below.

\subsection{Low Energy Gap Limit}
Let us consider the viability of setting the energy gap to be some arbitrary small number, not unlike a quantum system with degenerate eigenstates. Notice that \eqref{prob} and \eqref{corr} are both analytic functions of $\Omega$ within the domain $v\in[0,1]$ so we can simply set $\Omega=0$ and obtained a well defined limit. In practice, while computing Negativity, we do not need to worry about the value of energy gap despite being arbitrarily low. 
\begin{align} 
   &\lim_{\Omega\rightarrow 0} P=\frac{\lambda^2}{4\pi} 
   \label{zeroprob}
   \\
   & \lim_{\Omega\rightarrow 0} X =\lambda ^2\left(\frac{1-v^2}{8\pi i}\right)\int _{-\infty }^{\infty }du\,\,
   \frac{\text{Exp}\left[-(1-v^2)\frac{d^2+u^2(1+v^2)}{4\sigma
   ^2}\right]}{\sqrt{v^2u^2+d^2}}
   \nonumber
   \\
   &\qquad\qquad
   \times \left(1+\textbf{erf}\left[\frac{i   \sqrt{1-v^2}\sqrt{v^2u^2+d^2}}{2\sigma }\right]\right) 
   \nonumber
   \label{zerocorr}
\end{align}
The expression \eqref{zeroprob} is independent of switching width so that for any Gaussian switching function, it yields the same excitation probability. Furthermore it is trivial to verify that the expression in \eqref{zerocorr} is a well-defined limit. The only remaining parameters that affect Negativity in this limit are therefore the relative velocity $v$ and the ratio between distance and switching width. This is a property unique to $3+1$-dimensions, as equally scaling the distance and switching width does not yield equal Negativity in other dimensions (see Appendix A). 
\begin{figure}
    \centering
    \includegraphics[scale=0.6]{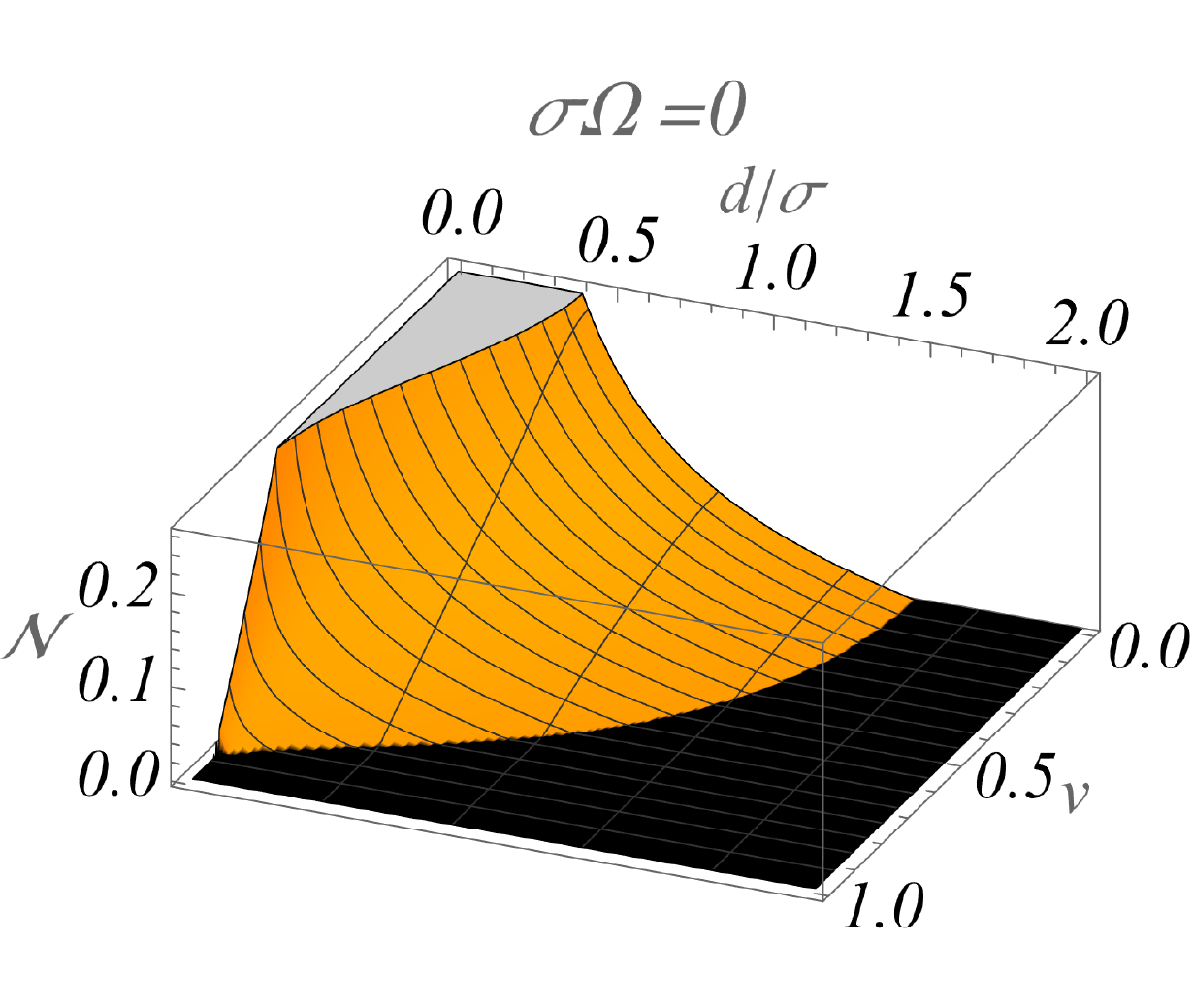}
    \caption{Plots of Negativity in units of $\lambda^2$ at $\sigma\Omega=0$ as a function of minimal distance $d$ and velocity $v$}
    \label{fig:zeroo}
\end{figure}
 In fig. \ref{fig:zeroo}, we can see that Negativity is indeed well defined and non-zero as $\Omega\rightarrow 0$ while still retaining the general properties: the Negativity increases as distance decreases while still retaining a maximum in $v$, as can be seen by the shifting colors. %The study of entanglement between two detectors with arbitrarily small energy gap is possible and follow the consistent procedure of finding a zero limit solution to the Negativity equation, expressed in \eqref{zeroprob} and \eqref{zerocorr}.  
 We also note from fig. \ref{fig:zeroo} that the detectors do not harvest entanglement when $d/\sigma\gtrsim 1.5$. This is consistent with the results of \cite{Pozas-Kerstjens:2017xjr},  which says that degenerate spacelike separated detectors cannot harvest entanglement.

\subsection{Optimization Condition}

\begin{figure}
    \centering
    \includegraphics[width=0.8\linewidth]{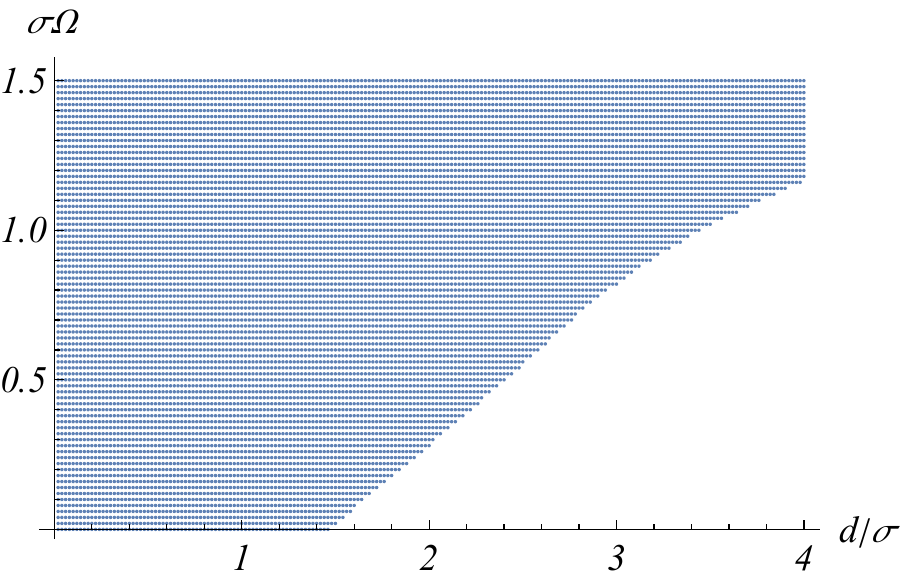}
    \caption{$d/\sigma$ vs values of $\sigma\Omega$. The blue dots approximate the region where entanglement can exist at some velocity} 
    \label{fig:optimization}
\end{figure}

\begin{figure}
    \centering
    \includegraphics[width=0.7\linewidth]{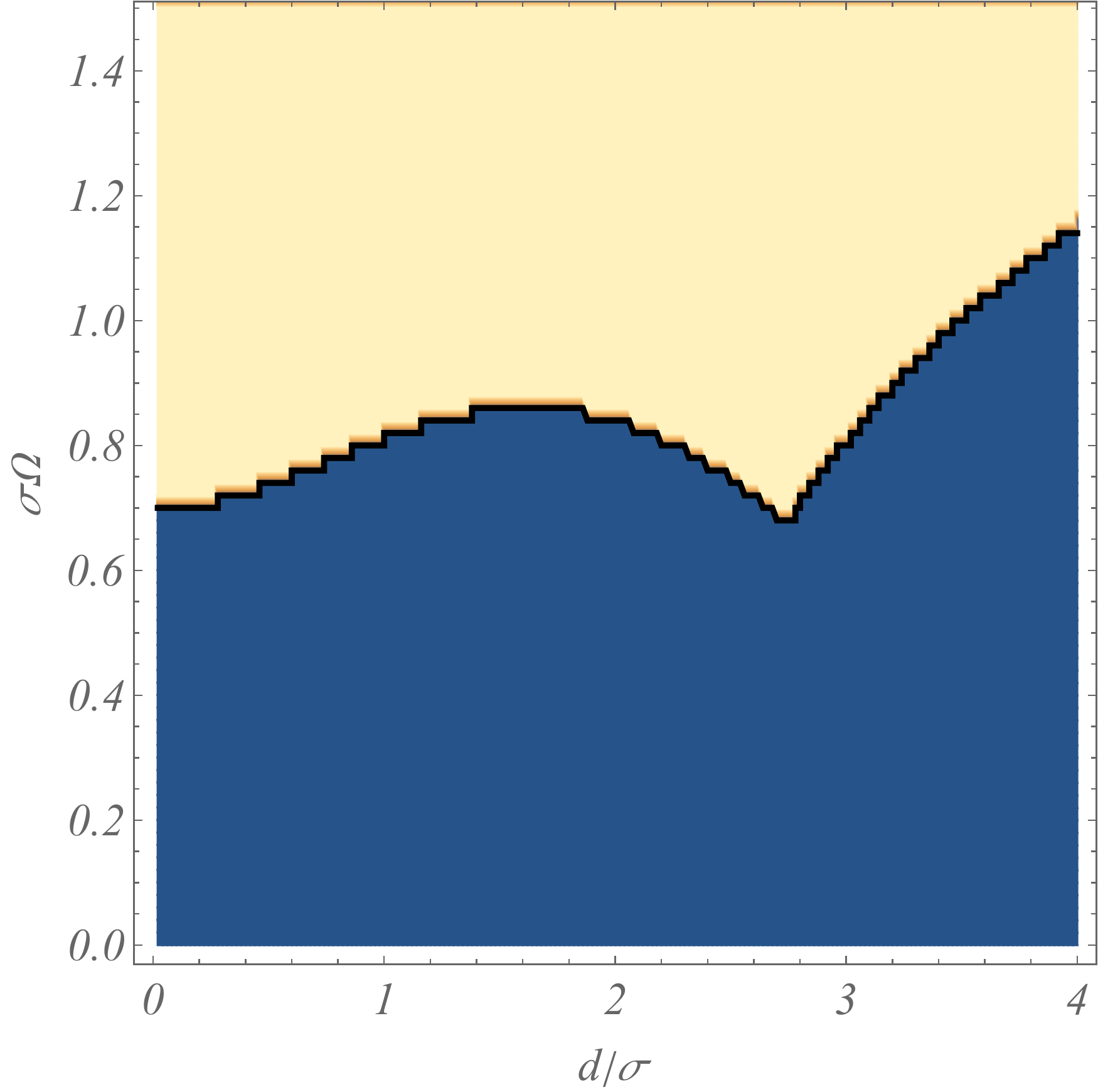}
    \caption{The cream-coloured region
 indicates $\mathcal{M}$ is initially increasing, implying there exists a peak in the Negativity.  In the blue region  $\mathcal{M}$ is non-increasing.    }
    \label{fig:possiblepeak}
\end{figure}

  We next consider how Negativity can be optimized over the parameters $\{v,d/\sigma,\sigma\Omega\}$. Our main interest is in the range of parameters that admit a non-trivial peak in $\mathcal{N}(v)$. It is impossible for $\mathcal{N}(v)$ to monotonically increase as it is a non-negative continuous function with value zero at lightspeed. Thus in the parameter region where there are no peaks in $\mathcal{N}(v)$, we either have $\mathcal{N}(v)$ equals zero throughout $v\in[0,1]$ or $\mathcal{N}(v)$ monotonically decreases as a function of velocity. As a truncated differentiable function, we can obtain this region by compartmentalizing it as two tasks: first, verify the existence of velocity $v_0>0$ such that the derivative of $d\mathcal{M}/dv|_{v=v_0}=0$, then check if $\mathcal{M}(v_0)>0$.
$\mathcal{N}(v)$ has thus far produced either one or zero critical points within $v\in(0,1)$ and therefore  we will work on the assumption that the continuous function $\mathcal{M}(v)$ can only obtain a local maximum in this range if it initially ``trends upward'' as velocity increases. Writing $\eta=v^2$, we can use the sign of the derivative $\frac{d\mathcal{M}}{d\eta}|_{\eta=0}$ to infer the relationship between parameters $\{d,\Omega\}$ and the growth trend of $\mathcal{M}$ at $v=0$.
From this (see Appendix B), we can obtain the energy gap $\Omega_p[d]$ that separates upward trending and downward trending $\mathcal{M}$ at distance $d$:
\begin{widetext}
\begin{equation}
    \Omega>\Omega_p[d]:=\frac{1}{2\sigma}\sqrt{2-\bigg(\frac{d}{\sigma}\bigg)^2+\frac{4\sqrt{\pi}\bigg(1+\textbf{Erfi}\big[\frac{d}{2\sigma}\big]^2\bigg)}{\sqrt{\pi}\bigg(1+\frac{2\sigma^2}{d^2}\bigg)\bigg(1+\textbf{Erfi}\big[\frac{d}{2\sigma}\big]^2\bigg)-\frac{2\sigma}{d}e^\frac{d^2}{4\sigma^2}\textbf{Erfi}\big[\frac{d}{2\sigma}\big]}}\,.
    \label{Omegaformula}
\end{equation}
\end{widetext}
We illustrate in fig. \ref{fig:optimization} the region where negativity can exist at some velocity. The region where there is a peak in the Negativity is shown in fig. \ref{fig:possiblepeak}, obtained by overlapping fig. \ref{fig:optimization} and the region corresponding to \eqref{Omegaformula}. 

\newpage

   Generally, at sufficiently large $\Omega$, increasing $\Omega$ suppresses $P$ to a higher degree than it does $|X|$, thus allowing for entanglement to survive at sufficiently high values of $\Omega$, albeit at ever diminishing amounts.
This is consistent with the static case as per fig. \ref{fig:my_label}. 
  We also observe also from  fig. \ref{fig:possiblepeak}   
  that after $d/\sigma$ exceeds
  $2.8$, the required gap to maintain entanglement increases monotonically with $d/\sigma$;  this trend continues for values of $d$ beyond the bounds in the figure. 
 The upper region of fig. \ref{fig:possiblepeak} is where  Negativity is initially increasing with
 respect to $v$. This in turn implies that there is a peak in the entanglement within this region
 of $(d,\Omega)$ parameter space.
 Note that sufficiently high energy gap allows for entanglement, and the necessary threshold for energy gap such that a peak exist increases as distance increases. 
 These are expected results based on previous discussions. We see that there exists no local maximum in $v\in(0,1)$ at values of $\sigma\Omega<0.065$.

\section{Conclusion}
\label{conclusion}

Our key result is that the 
Negativity harvested by a
pair of UDW detectors depends on  their relative velocity.
We have worked with identical detectors  
in a frame where they have equal and opposite velocity, and are offset in position by a separation $d$ at their point of closest approach, with
the peaks of their switching functions chosen to coincide at this point.  These limitations are chosen for simplicity and convenience; the general result that Negativity depends on relative velocity will be present even if the detectors are not identical and the peaks of their switching functions are offset.  

Our results
extend  the set of parameters available for optimising the entanglement harvested. Evidently, this detector configuration is able to harvest space-like entanglement at high values of energy gap, similar to the static case. The effect of minimal distance on Negativity is similar to the static case; namely a larger distance reduces the Negativity (unless it is previously zero). What is most surprising however, is that there exists a range of minimal distance and energy gap such that entanglement grows in comparison to the static case as the relative velocities increase. Similar effects have been previously observed in other contexts \cite{Koga:2018the,Barman:2022xht}, where  it was observed that the entanglement harvested, having a non-trivial dependence on $v$, vanishes in the limit $v\rightarrow 0 $, in agreement  with what we find\footnote{We thank B. R. Majhi for pointing out these interesting works to us.}. We also find that this region of growth in entanglement with velocity is $d$ and $\Omega$ dependent and exists at any distance with sufficiently high energy gap. 

Finally, we see that large energy gaps allow entanglement to survive at higher velocities and increases the ratio between the value of Negativity at zero relativity and the peak value. This behavior implies that detectors at large distances can become entangled  more effectively by moving at some suitable relative velocity, thus enhancing the parameter-space toolbox in information theory.

\begin{acknowledgments}
C.S. would like to thank Laura Henderson for help with the technical side of this paper, Rick Tales Perche for  advice on   writing  this paper, and Daniel Chen who provided constant reminders about doing research.
This work was supported in part by the Natural Sciences and Engineering Research Council of Canada.
\end{acknowledgments}

\appendix
\begin{widetext}
\section{Expressions for $P$ and $X$}

In this Appendix, we   calculate the excitation probability and correlation factor of detectors in $n$ dimensions through the use of mode-expansion. This is done by taking the Wightman function in its plane-wave form and integrating it with the corresponding time-associated variables. as the integrand is analytical, we are allowed to swap the integration order and avoid singularities through the use of polar coordinates

  To calculate $P$, we will begin by using the trajectory of detector B, which in all dimensions have been set such that it always have the displacement $v(t'-t)\hat{e}_1$ from time $t$ to $t'$. We expect that it to be Lorentz invariant, as typical of excitation probability over an infinite time period. If this is indeed the cases, then $P_D=P_A=P_B$, as $P_A$ is $P_B$ with its $v$ being substituted by its antiparallel counterpart. 

  The Wightman function in $n+1$ dimensional flatspace may be written as
\begin{equation}
    W(\textbf{x},\textbf{x}'):=\int\frac{d^n k}{2(2\pi)^n|k|}e^{i(|k|\Delta t-\vec{k}\cdot\vec{\Delta x)}}
\end{equation}
By (2.12) and using detector B trajectory, the excitation probability of scalar coupling detectors in $n+1>2$ dimensions is
\begin{align*}
    P_D&=\lambda^2\int\int d\tau d\tau' e^{-\tau^2/2\sigma^2}e^{-\tau'^2/2\sigma^2}e^{-i\Omega(\tau-\tau')}W(\textbf{x}_A(t),\textbf{x}_A(t'))\\
        &=\lambda^2\int\int d\tau d\tau' e^{-\tau^2/2\sigma^2}e^{-\tau'^2/2\sigma^2}e^{-i\Omega(\tau-\tau')}\int\frac{d^n k}{2(2\pi)^n|k|}e^{i\gamma(|k|(\tau'-\tau)-k_1v(\tau'-\tau) )}\\
         &=\lambda^2\int\frac{d^n k}{2(2\pi)^n|k|}\int\int d\tau d\tau' e^{-\tau^2/2\sigma^2}e^{-\tau'^2/2\sigma^2}e^{-i\Omega(\tau-\tau')}e^{i\gamma(|k|(\tau'-\tau)-k_1v(\tau'-\tau) )}
\end{align*}
As the integration domain is infinite, we can substitute $k$ with $k/\gamma$ without changing the limits of integration, and obtain
\begin{equation}
    P_D=\frac{\lambda^2\sigma^2}{2(2\gamma\pi)^{n-1}}\int\frac{d^n k}{|k|}e^{-(|k|+\Omega-v k_1)^2\sigma^2} \label{A.2}
\end{equation}

  In $n+1>2$ dimensions, the trajectory is such that $(x_{A,1}(t),x_{B,1}(t))=(-vt,vt)$ and $(x_{A,2}(t),x_{B,2}(t))=(-\frac{d}{2},\frac{d}{2})$. The other components of the trajectories are either constants or zero. Using this trajectory and inserting it in
  \eqref{Wightman}, we can generalize the correlation function in $n+1>2$ dimensions to
\begin{align*}
    X&=-\frac{1}{2}\lambda^2\int_{-\infty}^\infty du \,e^{-u^2/4\sigma^2} e^{-i\Omega u}\int_0^\infty ds\,e^{-s^2/4\sigma^2}\\& \bigg[W(\textbf{x}_A(\gamma(u-s)/2),\textbf{x}_B(\gamma(u+s)/2))+W(\textbf{x}_B(\gamma(u-s)/2),\textbf{x}_A(\gamma(u+s)/2))\bigg]\\
    &=-\frac{1}{2}\lambda^2\int_{-\infty}^\infty du \,e^{-u^2/4\sigma^2} e^{-i\Omega u}\int_0^\infty ds\,e^{-s^2/4\sigma^2}\\&
    \bigg[\int\frac{d^n k}{2(2\pi)^n|k|}e^{i\gamma(|k|s-v k_1 u-k_2d/\gamma)}+\int\frac{d^n k}{2(2\pi)^n|k|}e^{i\gamma(|k|s+v k_1 u+k_2d/\gamma)}\bigg]
\end{align*}
Notice that the two integrals are equal, as one of the integrals is simply the other one with $(k_1,k_2)$ substituted with $(-k_1,-k_2)$, with the inverted domain. Furthermore, as the integration domain of $k$ encompasses $\mathbb{R}^n$, we can make a change of variables such that $vk_1u+k_2d\to \sqrt{v^2u^2+d^2/\gamma^2}k_1$
\begin{align*}
    &=-\frac{1}{2}\lambda^2\int_{-\infty}^\infty du \,e^{-u^2/4\sigma^n} e^{-i\Omega u}\int_0^\infty ds\,e^{-s^2/4\sigma^2}\bigg[2\int\frac{d^2 k}{2(2\pi)^2|k|}e^{i\gamma(|k|s+\sqrt{v^2u^2+d^2/\gamma^2}k_1)}\bigg]\\
    &=-\frac{\lambda^2}{2(2\pi)^n}\int_{-\infty}^\infty du \,e^{-u^2/4\sigma^2} e^{-i\Omega u}\int\frac{d^n k}{|k|}e^{i\gamma\sqrt{v^2u^2+d^2/\gamma^2}k_1}\int_0^\infty ds\,e^{-s^2/4\sigma^2}e^{i\gamma|k|s}\\
    &=-\frac{\lambda^2}{2(2\pi)^n}\int_{-\infty}^\infty du \,e^{-u^2/4\sigma^2} e^{-i\Omega u}\int\frac{d^n k}{|k|}e^{i\gamma\sqrt{v^2u^2+d^2/\gamma^2}k_1}e^{-\gamma^2|k|^2\sigma^2}\sqrt{\pi}\sigma(1+i\,\textbf{Erfi}[\gamma|k|\sigma])\\
    &=-\frac{\lambda^2\sigma\sqrt{\pi}}{4\pi(2\pi)^{n-1}}\int_{-\infty}^\infty du \,e^{-u^2/4\sigma^2} e^{-i\Omega u}\int\frac{d^n k}{|k|}e^{i\gamma\sqrt{v^2u^2+d^2/\gamma^2}k_1}e^{-\gamma^2|k|^2\sigma^2}(1+i\,\textbf{Erfi}[\gamma|k|\sigma])
\end{align*}
Again, substituting $k \to k/\gamma$, we obtain
\begin{equation}
    \hspace{-1cm}X=-\frac{\lambda^2\sigma\sqrt{\pi}}{4\pi(2\pi\gamma)^{n-1}}\int_{-\infty}^\infty du \,e^{-u^2/4\sigma^2} e^{-i\Omega u}\int\frac{d^n k}{|k|}e^{i\sqrt{v^2u^2+d^2/\gamma^2}k_1}e^{-|k|^2\sigma^2}(1+i\,\textbf{Erfi}[|k|\sigma]) 
\end{equation}\\

\subsection{3+1 Dimensions}

Setting $k_1=r \cos{\theta}$, $k_2=r\sin{\theta}\cos{\phi}$, and $k_3=r\sin{\theta}\sin{\phi}$, in $(3+1)$ dimensions 
we obtain
\begin{align*}
    P&=\frac{\lambda^2\sigma^2}{8\pi^2\gamma^2}\int\frac{d^3 k}{|k|}e^{-(|k|+\Omega- v k_1)^2\sigma^2} 
    =\frac{\lambda^2\sigma^2}{8\pi^2\gamma^2}\int_0^\infty\int_0^\pi\int_0^{2\pi}\frac{r^2\sin{\theta}\,dr\,d\theta\,d\phi}{r}e^{-(r+\Omega- v r\cos{\theta})^2\sigma^2}\\
    &=\frac{\lambda^2\sigma^2}{4\pi\gamma^2}\int_0^\pi \sin{\theta}\,d\theta\bigg(\frac{e^{-\sigma^2\Omega^2}-\sqrt{\pi}\sigma\Omega\,\textbf{Erfc}[\sigma\Omega]}{2\sigma^2(1-v\cos{\theta})^2}\bigg)\\
    &=\frac{\lambda^2\sigma^2}{4\pi\gamma^2}\bigg(\frac{e^{-\sigma^2\Omega^2}-\sqrt{\pi}\sigma\Omega\,\textbf{Erfc}[\sigma\Omega]}{\sigma^2(1-v^2)}\bigg)\\
    &=\frac{\lambda^2}{4\pi}\bigg(e^{-\sigma^2\Omega^2}-\sqrt{\pi}\sigma\Omega\,\textbf{Erfc}[\sigma\Omega]\bigg)
\end{align*}
which is clearly Lorentz Invariant. 
Likewise we obtain
\begin{align*}
    X&= -\frac{\lambda^2\sigma\sqrt{\pi}}{4\pi(2\pi\gamma)^2}\int_{-\infty}^\infty du \,e^{-u^2/4\sigma^2} e^{-i\Omega u}\int\frac{d^3 k}{|k|}e^{i\sqrt{v^2u^2+d^2}k_1}e^{-|k|^2\sigma^2}(1+i\,\textbf{Erfi}[|k|\sigma]);\\
    &=-\frac{\lambda^2\sigma\sqrt{\pi}}{16\pi^3\gamma^2}\int_{-\infty}^\infty du \,e^{-u^2/4\sigma^2} e^{-i\Omega u}\int_0^\infty\int_0^\pi\int_0^{2\pi}\frac{r^2\sin{\theta}\,dr\,d\theta\,d\phi}{r}e^{i\sqrt{v^2u^2+d^2/\gamma^2}r\cos{\theta}}e^{-r^2\sigma^2}(1+i\,\textbf{Erfi}[r\sigma])\\
    &=-\frac{\lambda^2\sigma\sqrt{\pi}}{8\pi^2\gamma^2}\int_{-\infty}^\infty du \,e^{-u^2/4\sigma^2} e^{-i\Omega u}\int_0^\infty re^{-r^2\sigma^2}(1+i\,\textbf{Erfi}[r\sigma])\,dr\int_0^\pi \sin{\theta}\,d\theta e^{i\sqrt{v^2u^2+d^2/\gamma^2}r\cos{\theta}}\\
     &=-\frac{\lambda^2\sigma\sqrt{\pi}}{4\pi^2\gamma^2}\int_{-\infty}^\infty du \frac{1}{\sqrt{u^2v^2+d^2/\gamma^2}}\,e^{-u^2/4\sigma^2} e^{-i\Omega u}\int_0^\infty e^{-r^2\sigma^2}(1+i\,\textbf{Erfi}[r\sigma])\sin{(r\sqrt{u^2v^2+d^2/\gamma^2})}\,dr\\
     &=-\frac{\lambda^2\sigma\sqrt{\pi}}{4\pi^2\gamma^2}\int_{-\infty}^\infty du \frac{1}{\sqrt{d^2/\gamma+u^2v^2}}\,e^{-u^2/4\sigma^2} e^{-i\Omega u}\bigg(e^{-\frac{d^2/\gamma^2+u^2v^2}{4\sigma^2}}\frac{\sqrt{\pi}}{2\sigma}\bigg(i+\textbf{Erfi}\bigg[\frac{\sqrt{d^2/\gamma^2+u^2v^2}}{2\sigma}\bigg]\bigg)\bigg)\\
    &=\lambda ^2\left(\frac{1-v^2}{8\pi i}\right)\int _{-\infty }^{\infty }du\,\,
   \frac{\text{Exp}\left[-\frac{d^2(1-v^2)+u^2(1-v^4)}{4\sigma
   ^2}\right]}{\sqrt{v^2u^2+d^2}} e^{-i\Omega u\sqrt{1-v^2}}\left(1+\textbf{Erf}\left[\frac{i  \sqrt{1-v^2}\sqrt{v^2u^2+d^2}}{2\sigma }\right]\right) 
\end{align*}
upon  substituting $u\to u/\gamma$.
\section{Derivation of $\Omega_p[d]$}

  By Appendix (A.1), the correlation is
 
\begin{align*}
    X&=\frac{\lambda^2}{8\pi i\gamma^2}\int_{-\infty}^\infty du \frac{1}{\sqrt{d^2/\gamma^2+u^2v^2}}\,e^{-\frac{d^2/\gamma^2+u^2(1+v^2)}{4\sigma^2}} e^{-i\Omega u}\bigg(1-i\,\textbf{Erfi}\bigg[\frac{\sqrt{d^2/\gamma^2+u^2v^2}}{2\sigma}\bigg]\bigg)\\
  &=\frac{\lambda^2}{8\pi i\gamma^2}\frac{e^{-\frac{d^2}{4\sigma^2}}e^{-i\Omega\sigma}}{2d^3\sigma}\Bigg[4 \sqrt{\pi } d^2 \sigma ^2 \left(1-i \textbf{Erfi}\left[\frac{d}{2 \sigma }\right]\right)+
  \\& \left(\sqrt{\pi } \left(1-i \textbf{Erfi}\left[\frac{d}{2 \sigma }\right]\right) \left(d^2+2 \sigma ^2\right)+2 i d \sigma  e^{\frac{d^2}{4 \sigma ^2}}\right)\left(d^2-2 \sigma ^2+4 \sigma ^4 \Omega ^2\right)v^2+\mathcal{O}\left(v^4\right)\Bigg]\,.
\end{align*}
Hence,
\begin{align*}
    \frac{d|X|^2}{dv^2}\Bigg|_{v=0}&=e^{-2 \sigma ^2 \Omega ^2} \Bigg(\pi  e^{-\frac{d^2}{2 \sigma ^2}} \left(\textbf{Erfi}\left[\frac{d}{2 \sigma }\right]^2+1\right) \left(d^4+4 d^2 \sigma ^2 \left(\sigma ^2 \Omega ^2-1\right)+8 \sigma ^6 \Omega ^2-4 \sigma ^4\right)
    \\&-4 d \sigma  \textbf{DawsonF}\left[\frac{d}{2 \sigma }\right] \left(d^2+4 \sigma ^4 \Omega ^2-2 \sigma ^2\right)\Bigg)/32 \pi ^2 d^4
    \\\frac{d|X|^2}{dv^2}>0&\iff \Omega>\Omega_p[d]:=\frac{1}{2\sigma}\sqrt{2-\bigg(\frac{d}{\sigma}\bigg)^2+\frac{4\sqrt{\pi}\bigg(1+\textbf{Erfi}\big[\frac{d}{2\sigma}\big]^2\bigg)}{\sqrt{\pi}\bigg(1+\frac{2\sigma^2}{d^2}\bigg)\bigg(1+\textbf{Erfi}\big[\frac{d}{2\sigma}\big]^2\bigg)-\frac{2\sigma}{d}e^\frac{d^2}{4\sigma^2}\textbf{Erfi}\big[\frac{d}{2\sigma}\big]}}
    \\&\frac{d|X|^2}{dv^2}\Bigg|_{v=0}>0\iff \frac{d|X|}{dv^2}>0\text{, since $|X|_{v=0}>0$}
\end{align*}
It is then evident   that $|X|$, and by extension Negativity, only has an upward trend if and only if $\Omega>\Omega_p[d]$.
\end{widetext}

\bibliography{EHrelativeVsubmit}% Produces the bibliography via BibTeX.

\end{document}